\documentclass[twocolumn,showpacs,preprintnumbers,amsmath,amssymb]{revtex4-1}
\usepackage{graphicx}
\usepackage{textcomp}

\begin{document}
\title{Optical realization of two-boson tunneling dynamics}
  \normalsize
\author{Stefano Longhi 
}
\address{Dipartimento di Fisica, Politecnico di Milano, Piazza L. da Vinci
32, I-20133 Milano, Italy}

%
\bigskip
\begin{abstract}
An optical realization of the tunneling dynamics of two interacting
bosons in a double-well potential, based on light transport in a
four-core microstructured fiber, is proposed. The optical setting
enables to visualize in a purely classical system the entire
crossover from Rabi oscillations to correlated pair tunneling and to
tunneling of a fragmented pair in the fermionization limit.
 \noindent
\end{abstract}



\maketitle

\section{Introduction.}
  Light transport in engineered optical waveguides has
provided a fascinating and experimentally accessible framework to
visualize in a classical setting many universal coherent quantum
phenomena generally encountered in condensed-matter or matter-wave
systems \cite{Christodoulides03,Longhi09}. This has lead to the
prediction and observation of a wide variety of classic optics
analogues of {\em single-particle} nonrelativistic and even
relativistic phenomena, such as Bloch oscillations and Zener
tunneling \cite{Christodoulides03,BO}, dynamic localization
\cite{Longhi06}, Anderson localization \cite{and}, coherent
destruction of tunneling \cite{Longhi09}, Zeno dynamics \cite{Zeno},
adiabatic stabilization \cite{Longhi05}, and Zitterbewegung
\cite{Dreisow10}. Since photons do not interact, it is a common
belief that, as opposed to other quantum systems such as cold atoms
or trapped ions (see, e.g., \cite{JPBreview}), the use of photonics
as a model system for quantum physics carries the intrinsic drawback
of being limited to visualize {\em single-particle} phenomena,
missing the possibility to simulate the richer physics of
interacting many-particle quantum systems. A paradigmatic example of
many-body physics is found in quantum tunneling of bosons in a
double well potential, the so-called bosonic junction
\cite{C0,C1,C2,C3,C4,C5,C6}. For a relatively large number and
weakly interacting bosons,  this has led to the observation of
Josephson oscillations and nonlinear self-trapping of bosons above a
critical interaction strength, as described by a standard
Bose-Hubbard model or by coupled mean-field equations in the
Gross-Pitaevskii limit \cite{C0}. A simple optical realization of
the bosonic junction in such a limiting case is based on light
tunneling between two coupled nonlinear waveguides \cite{Ruffo}.
However, a richer dynamical scenario has been recently predicted to
occur for tunneling of {\em few} and {\em strongly} correlated
bosons \cite{C4,vari}, covering the full crossover from weak
interactions to the fermionization limit of the Tonks-Girardeau gas
\cite{fermionization}. In particular, the tunneling dynamics of two
bosons in a one-dimensional double well shows a transition from Rabi
oscillations, in the absence of interaction, to correlated pair
tunneling and further to fragmented-pair tunneling as the
interaction strength is increased \cite{C4}. As few-body
counterparts of the self-trapping transition and correlated pair
tunneling in a bosonic junction have been reported in recent
experiments \cite{C3,C7}, an observation of the rich two-boson
tunneling dynamics up to the fermionization limit \cite{C4} is still
missing. In this article it is shown that such a two-boson tunneling
dynamics can be realized in a classical optical setting based on
four-core guiding dielectric structure, in which the electric field
propagation along the guide mimics the quantum mechanical evolution
of the two-particle wave function. The paper is organized as
follows. In Sec.II, the quantum-optical analogy between light
propagation in a four-core microstructured fiber and the dynamics of
two interacting bosons in a double well is outlined. In Sec.III, a
detailed analysis of the tunneling dynamics is presented, and the
entire crossover from Rabi oscillations to correlated pair tunneling
and to tunneling of a fragmented pair in the fermionization limit is
explained on the basis of the coupling among the various modes
sustained by the fiber cores. Finally, in Sec.IV the main
conclusions are outlined.

\section{Quantum-optical analogy}
Let us consider a weakly-guiding dielectric structure with a
refractive index $n(x_1,x_2)$, which varies in the transverse
$(x_1,x_2)$ plane but remains invariant along the axial direction
$z$. In the paraxial approximation, propagation of monochromatic
light waves is described by a Sch\"{o}dinger-type wave equation for
the electric field envelope $\psi$  \cite{Longhi09}
\begin{equation}
i \lambdabar \partial_z \psi=-\frac{\lambdabar^2}{2 n_s} \left(
\frac{\partial^2}{\partial x_1^2}+ \frac{\partial^2}{\partial x_2^2}
\right) \psi+V(x_1,x_2) \psi,
\end{equation}
\begin{figure}
\includegraphics[scale=0.42]{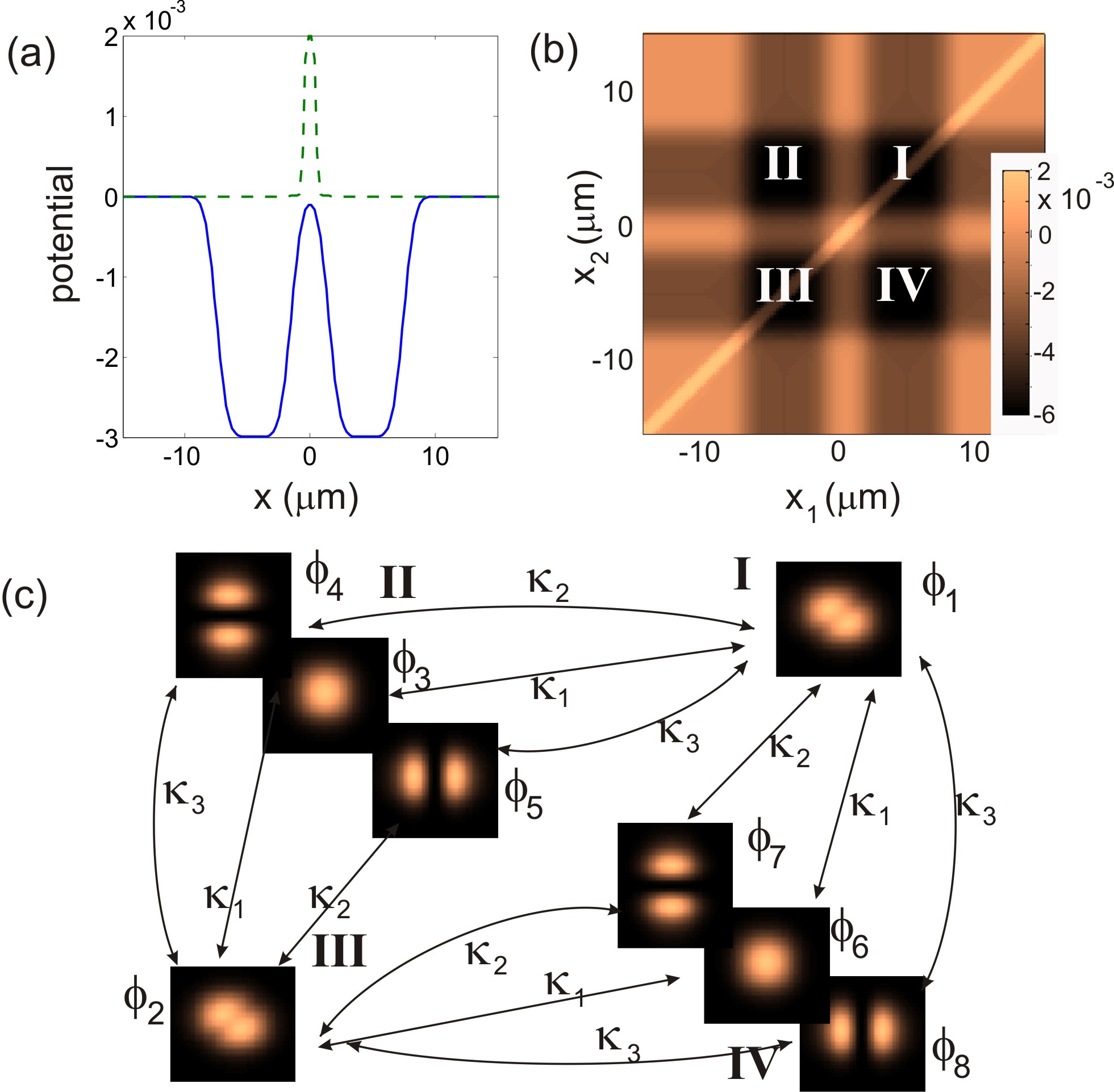}
\caption{ (color online)  (a) Profiles of the double well potential
$V_w(x)$ (solid curve) and of the repulsive potential $V_{int}(x)$
(dashed curve) for parameter values $a=4.5 \; \mu$m, $w=3 \; \mu$m,
$D_x=1 \; \mu$m, $w_i=0.5 \; \mu$m, $D_{xi}=0.2 \; \mu$m, $\Delta
n_1=0.003$, and $\Delta n_2=0.002$. The corresponding
two-dimensional optical potential $V(x_1,x_2)$ is shown in (b). (c)
Schematic of the guided modes supported by the four core regions
involved in the tunneling dynamics and their couplings.}
\end{figure}
where $\lambdabar=\lambda / (2 \pi)$ is the reduced wavelength of
photons, $V(x_1,x_2) \simeq n_s-n(x_1,x_2)$ is the optical
potential, and $n_s$ is the substrate refractive index. The
normalization condition $\int_{-\infty}^{\infty} dx_1 dx_2
|\psi|^2=1$ will be assumed in the following. Previous
quantum-optical analogies have generally viewed the paraxial wave
equation (1) as formally equivalent to the Schr\"{o}dinger equation
for a {\em single} particle of mass $n_s$ in a two-dimensional
potential $V(x_1,x_2)$, in which the temporal evolution of the
quantum particle is mapped into the spatial light evolution along
the axial direction $z$ and the Planck's constant is replaced by the
reduced wavelength of photons (see, for instance, \cite{Longhi05}).
However, whenever the potential $V$ has the form
\begin{equation}
V(x_1,x_2)=V_w(x_1)+V_w(x_2)+V_{int}(|x_1-x_2|),
\end{equation}
\begin{figure}
\includegraphics[scale=0.43]{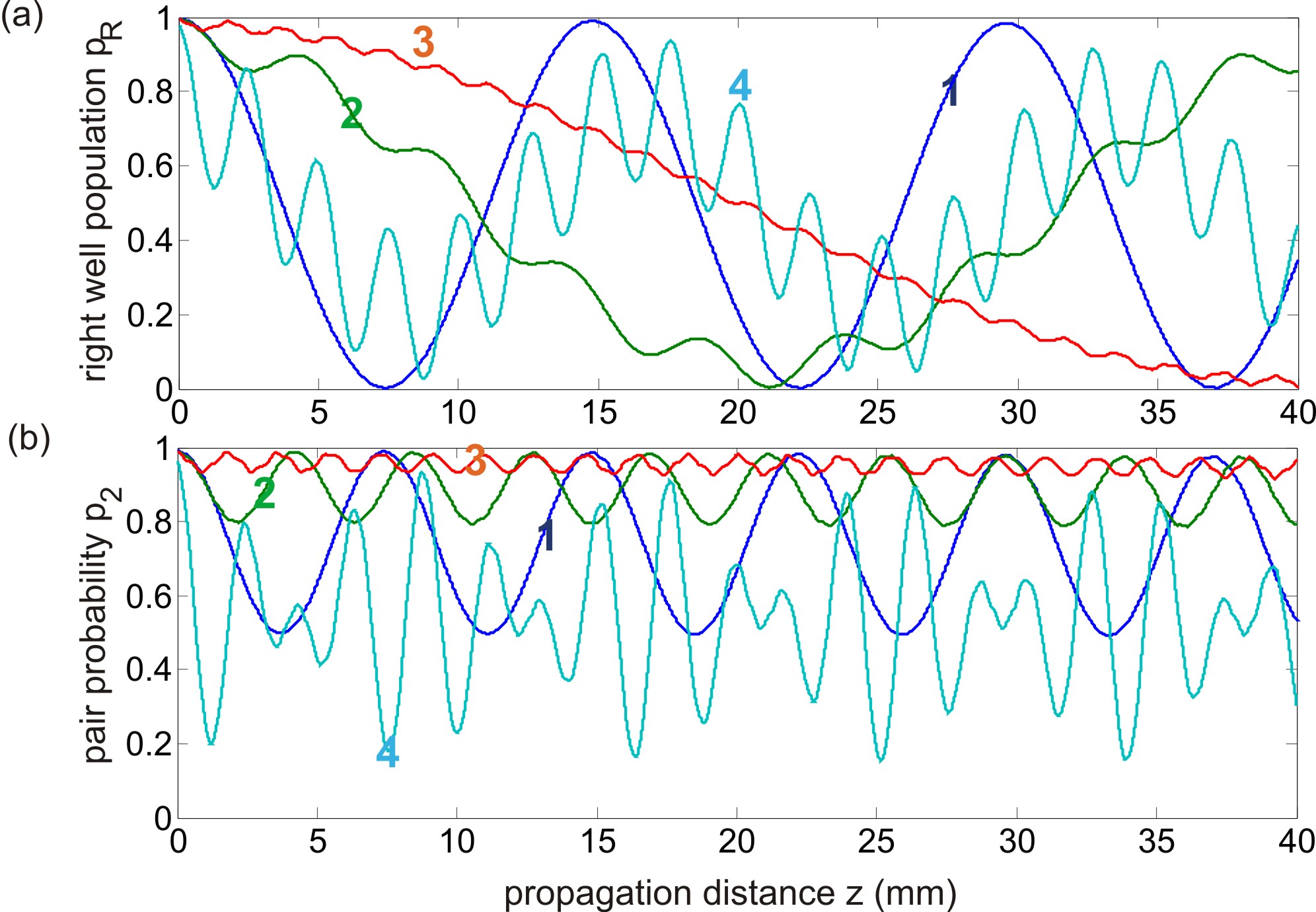}
\caption{ (color online)  Numerically-computed behavior of (a) the
percentage of bosons in the right well $p_R$, and (b) of the pair
probability $p_2$ versus propagation distance, for increasing values
of the particle interaction strength, measured by the index change
$\Delta n_2$. Curve 1: $\Delta n_2=0$ (non-interacting bosons);
curve 2:  $\Delta n_2=0.5 \times 10^{-3}$; curve 3:  $\Delta n_2=1.5
\times 10^{-3}$; curve 4:  $\Delta n_2=15 \times 10^{-3}$. Curves 2
and 3 correspond to the correlated pair tunneling regime, whereas
curve 4 corresponds to tunneling of a fragmented pair.}
\end{figure}
 where $V_w(x)$ is an arbitrary one-dimensional potential and $V_{int}(x)$ is a short-range
potential, Eq.(1) can be regarded as the optical analogue of the
Schr\"{o}dinger equation for {\em two} particles with the same mass
$n_s$ in a one-dimensional potential $V_w$, which interacts via the
potential $V_{int}$.  If the optical structure is excited at the
$z=0$ input plane by a beam satisfying the symmetry constraint
$\psi(x_1,x_2,0)=\psi(x_2,x_1,0)$, the wave function $\psi$ remains
symmetric along the propagation, and Eq.(1) thus describes the
evolution of two interacting identical bosons. Therefore, if we
assume for $V_w$ a double well shape and for $V_{int}$ a short-range
repulsive potential, our optical system realizes a classic wave
optics analogue of the two-boson junction recently studied in
Ref.\cite{C4}. In our optical system, we assume for $V_w(x)$ a
double well of the form \cite{DellaValle07} $V_{w}=-\Delta n_1
[g(x-a)+g(x+a)]$, where $g(x)=[{\rm erf}((x+w)/D_x)-{\rm
erf}((x-w)/D_x)]/[2 {\rm erf}(w/D_x)]$ is the well shape, $2a$ is
the distance between the two wells, $\Delta n_1>0$ is the peak index
change that defines the well depth, and $2w$ is the well width. For
the repulsive potential, we assume a similar functional form
$V_{int}=\Delta n_2 [{\rm erf}((x+w_i)/D_{xi})-{\rm
erf}((x-w_i)/D_{xi})]/[2 {\rm erf}(w_i/D_{xi})]$, with $w_i$ and
$D_{xi}$ much smaller than $w$ and $D_x$, respectively. The
refractive index change $\Delta n_2>0$ measures the strength of the
interaction, $\Delta n_2=0$ corresponding to non-interacting bosons.
Typical shapes of $V_w(x)$, $V_{int}(x)$ and of the resulting
two-dimensional potential $V(x_1,x_2)$ [Eq.(2)] are shown in
Figs.1(a) and (b). Note that the resulting potential $V$ in the
$(x_1,x_2)$ plane defines four higher-index guiding regions, i.e.
four waveguides, denoted by I-IV in Fig.1(b), which are evanescently
coupled. Such a four-core guide could be realized, for example, with
the technology of microstructured fibers \cite{fibers}, in which a
preform with the desired geometrical and refractive index features
is first manufactured. For example, using a cladding region made of
fused silica, the structure of Fig.1(b) could be realized by
assembling different regions of fused silica with different GeO$_2$
doping concentrations.

\section{Tunneling dynamics}
 The main features of the tunneling dynamics of two bosons in a double-well potential
 are captured by analyzing the evolution of the percentage of bosons in the right
well $p_R(z)$ and the the pair (or same-site) boson probability
$p_2(z)$, which are defined by \cite{C4}
\begin{eqnarray}
 p_R(z) & = & \int_{0}^{\infty} dx_1
\int_{-\infty}^{\infty}dx_2 |\psi|^2 \\
p_2(z) & = & \int_{x_1,x_2>0} dx_1 dx_2
 |\psi|^2+\int_{x_1,x_2<0} dx_1 dx_2 |\psi|^2 \;\;\;\;\;\;\;.
\end{eqnarray}
In our optical setting, $p_R(z)$ and $p_2(z)$  simply correspond to
the fractional light power trapped in waveguides I and IV, and in
waveguides I and III, respectively. A typical evolution of $p_R(z)$
and $p_2(z)$, as obtained by numerical integration of Eq.(1) for
increasing values of the interaction strength $\Delta n_2$, is shown
in Fig.2. Parameter values used in simulations are $\lambda=633$ nm
and $n_s=1.45$. In each simulation, the structure is excited at
$z=0$ in the fundamental mode of the guide I, which corresponds to
have initially the two bosons in the right-side well in the lowest
energy state. The scenario shown in Fig.2 reproduces the transition
from uncorrelated tunneling to pair tunneling and fragmented
tunneling in the fermionization limit, predicted in Ref.\cite{C4}.
For non-interacting bosons (curve 1), the atoms simply Rabi
oscillate back and forth between both wells, and they tunnel
independently. As a small correlation is introduced (curve 2), both
atoms tend to remain in the same well in the course of tunneling,
i.e they tunnel {\em as pairs}. Such a dynamical behavior, which was
observed in \cite{C5} and referred to as second-order tunneling, can
be simply explained in the framework of a standard two-site
Bose-Hubbard model, the optical simulation of which was recently
proposed in the Fock space using waveguide arrays \cite{Longhiun}.
However, the standard Bose-Hubbard model fails to predict the
tunneling regimes at strong interaction and the transition to the
fermionization limit. Indeed, at a larger interaction (curve 3),
tunneling tends to be inhibited, which is the few-body signature of
the self-trapping phenomenon of many bosons in the mean-field limit.
Remarkably, at stronger interaction and near the fermionization
limit (curve 4), tunneling is again allowed, and a fast oscillation
of $p_R(z)$ is superimposed to the slower tunneling cycle. This
basically corresponds to fragmented-pair tunneling at the Rabi
frequency predicted in Ref.\cite{C4}. Correspondingly, $p_2(z)$
passes through just about any value from 1 (fragmented pair) to
small values (near complete isolation). A detailed explanation of
such a rich tunneling scenario requires an inspection of the
low-lying energy spectrum of the exact two-boson Hamiltonian (1)
beyond the standard two-mode Bose-Hubbard approximation \cite{C4}.
In the optical context, the scenario can be explained in a different
view as the result of evanescent photonic tunneling among {\em a
few} guided modes of the {\em four} two-dimensional guides in the
geometrical setting of Fig.1(b). In fact, let us indicate by
$\phi_{1,2}$ the fundamental modes of the isolated waveguides I and
III, by $\phi_{3,4,5}$ the fundamental and the two lowest
higher-order degenerate transverse modes of the isolated waveguide
II, and by $\phi_{6,7,8}$ the fundamental and the two lowest
higher-order degenerate transverse modes of the isolated waveguide
IV. A typical profile of such modes is shown in Fig.1(c). Let then
expand the envelope $\psi$ as a superposition of such modes with
$z$-varying coefficients, i.e. $\psi=\sum_{l=1}^{8}c_l(z)
\phi_l(x_1,x_2) \exp(i \beta z)$, where $\beta$ is a reference
propagation constant. Note that, for symmetry reasons, one has
$c_6=c_3$, $c_7=c_4$ and $c_8=c_5$. In the tight-binding and
nearest-neighboring approximation, neglecting cross-coupling terms,
the following coupled-mode equations for the amplitudes $c_l$ can be
derived (see, for instance, \cite{Christodoulides03,Ruffo}):
\begin{eqnarray}
i (d c_1/dz) & = &  -2 \kappa_1 c_3-2 \kappa_2 c_4-2 \kappa_3
c_5+\delta_1 c_1 \nonumber
\\
i (d c_2/dz) & = &  -2 \kappa_1 c_3-2 \kappa_3 c_4-2 \kappa_2 c_5+
\delta_1 c_2 \nonumber
\\
i (d c_3/dz) & = &  -\kappa_1 (c_1+c_2)
\\
i (d c_4/dz) & = &  -\kappa_2 c_1- \kappa_3 c_2+ \delta_2 c_4
\nonumber
\\
i (d c_5/dz) & = &  -\kappa_3 c_1- \kappa_2 c_2+ \delta_2 c_5
\nonumber
\end{eqnarray}
\begin{figure}
\includegraphics[scale=0.43]{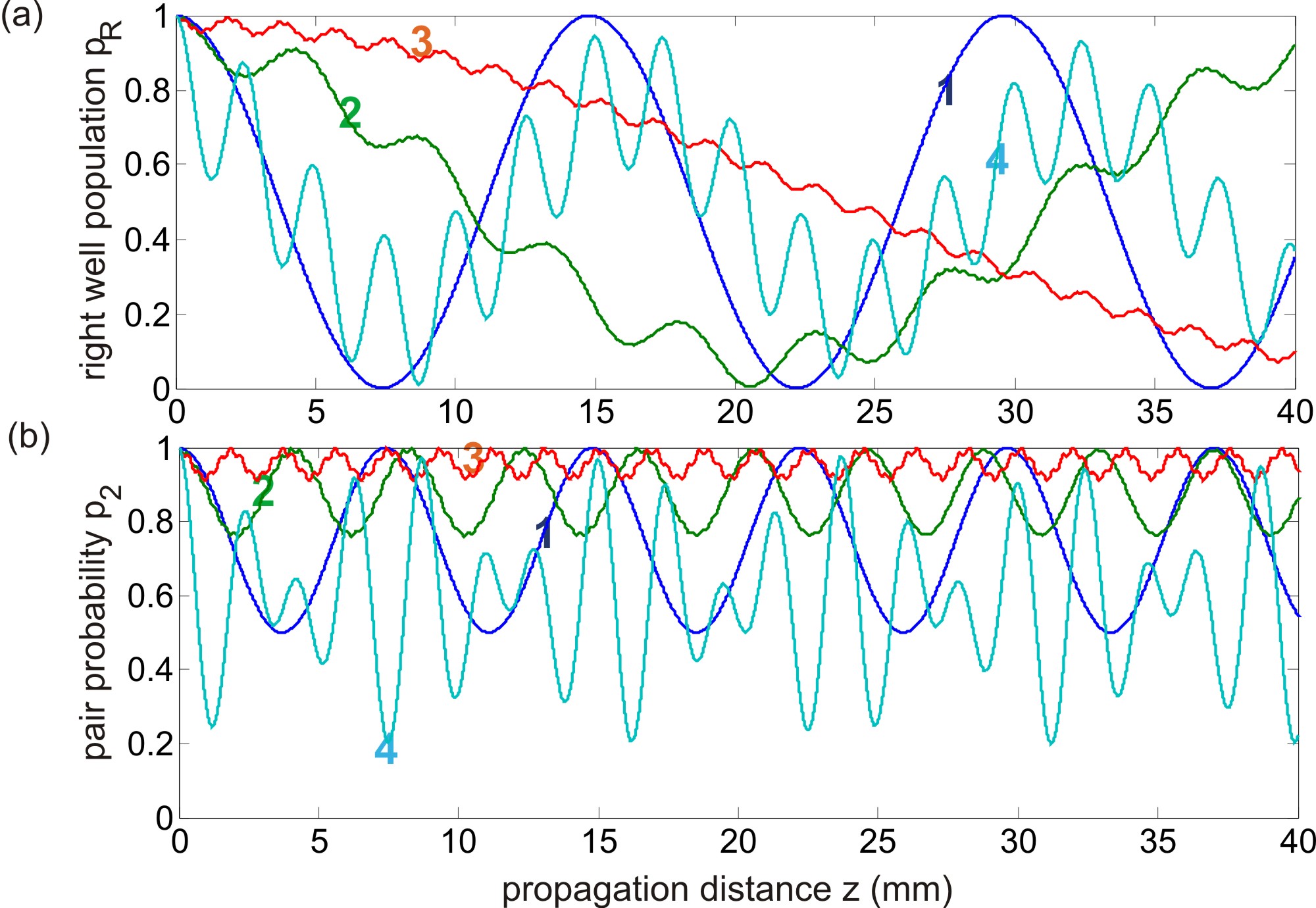}
\caption{ (color online)  (a) Behavior of $p_R$, and (b) of $p_2$
versus propagation distance, for increasing values of particle
interaction strength measured by the detuning $\delta_1$, as
predicted by the coupled-mode equations (3), for $\kappa_2=0.16$,
$\kappa_3=0.80$, and $\delta_2=20$ (in units of ${\rm mm}^{-1}$).
Curve 1: $\delta_1=0$, $\kappa_1=0.212$; curve 2: $\delta_1=1.22$,
$\kappa_1=0.26$; curve 3: $\delta_1=3.2$, $\kappa_1=0.32$; curve 4:
$\delta_1=18.9$, $\kappa_1=0.38$.}
\end{figure}
where $\kappa_1$, $\kappa_2$ and $\kappa_3$ are the coupling
constants between the couples of modes $\{\phi_1, \phi_3\}$,
$\{\phi_1,\phi_4\}$ and  $\{\phi_1,\phi_5\}$, respectively [see
Fig.1(c)], $\delta_1=\beta_3-\beta_1$ is the mismatch between the
propagation constants $\beta_3$ and $\beta_1$ of modes $\phi_3$ and
$\phi_1$, and $\delta_2=\beta_3-\beta_4$ is the mismatch between the
propagation constants $\beta_3$ and $\beta_4$ of modes $\phi_3$ and
$\phi_4$ (or $\phi_5$). Initial condition for Eqs.(5) is
$c_l(0)=\delta_{l,1}$. In terms of the amplitudes $c_l$, the
percentage of bosons in the right well and the same-site boson
probability, as defined by Eqs.(3) and (4), take take the simple
form
\begin{eqnarray}
p_R(z) & = & |c_1|^2+|c_3|^2+|c_4|^2+|c_5|^2 \\
p_2(z) & = & |c_1|^2+|c_2|^2,
\end{eqnarray}
respectively. In the absence of interaction, i.e. for $\Delta
n_2=0$, one has $\delta_1=0$ whereas $\delta_2$ is much larger than
the coupling constants. Hence, the higher-order transverse modes of
waveguides II and IV are not excited, i.e. one has $c_4 \sim c_5
\sim 0$, and the evolution of $c_1$, $c_2$ and $c_3$ can be
calculated exactly, yielding $p_R(z)= \cos^2(\kappa_1z)$ and
$p_2(z)=(1/2)[1+\cos^2(2 \kappa_1 z)]$: this is precisely the
dynamical behavior of uncorrelated bosons (curve 1 of Fig.2). As the
interaction $\Delta n_2$ is increased, the detuning $\delta_1$
increases, whereas $\delta_2$ does not change. The coupling
constants $\kappa_1$, $\kappa_2$ and $\kappa_3$ are given by
overlapping integrals involving the coupled
 guided modes, and are expected to slightly increase
as $\Delta n_2$ is increased because of the less confinement of
modes $\phi_1$ and $\phi_2$. If $(\delta_2-\delta_1)$ is still large
enough that the higher-order transverse modes of waveguides II and
IV are still out of resonance, the amplitudes $c_4$ and $c_5$ remain
small, and the tunneling dynamics is mainly governed by the first
three equations of the system (3), but with $\delta_1 \neq 0$. A
nonvanishing value of the detuning $\delta_1$ is responsible for the
doubly-periodicity of $p_R(z)$, the increase of the tunneling
period, and the appearance of correlated pair tunneling (i.e.
$p_2(z) \simeq 1$) as observed in curves 2 and 3 of Fig.2. As the
interaction $\Delta n_2$ is further increased, excitation of the
higher-order transverse modes of waveguides II and IV can not be
anymore neglected, and the tunneling dynamics requires to account
for the full five amplitudes entering in Eq.(5). For very strong
interactions, corresponding to the fermionization limit, the
fundamental modes $\phi_{1,2}$ of waveguides I and III get close to
resonance with the (degenerate) transverse modes $\phi_{4,5}$ and
$\phi_{7,8}$ of waveguides II and IV, whereas their fundamental
modes $\phi_{3,6}$ are now out of resonance. Hence, in the
fermionization limit one can set $c_3 \simeq 0$ in Eqs.(5).  Such
equations well describe the restoration of tunneling of a fragmented
pair. \par
A typical dynamical evolution of $p_{R}(z)$ and $p_{2}(z)$
in the various parameter regions, as obtained by numerically solving
the coupled-mode equations (5) by varying $\delta_1$ and taking into
account for the correction of $\kappa_1$ solely, is shown in Fig.3.
\begin{figure}
\includegraphics[scale=0.44]{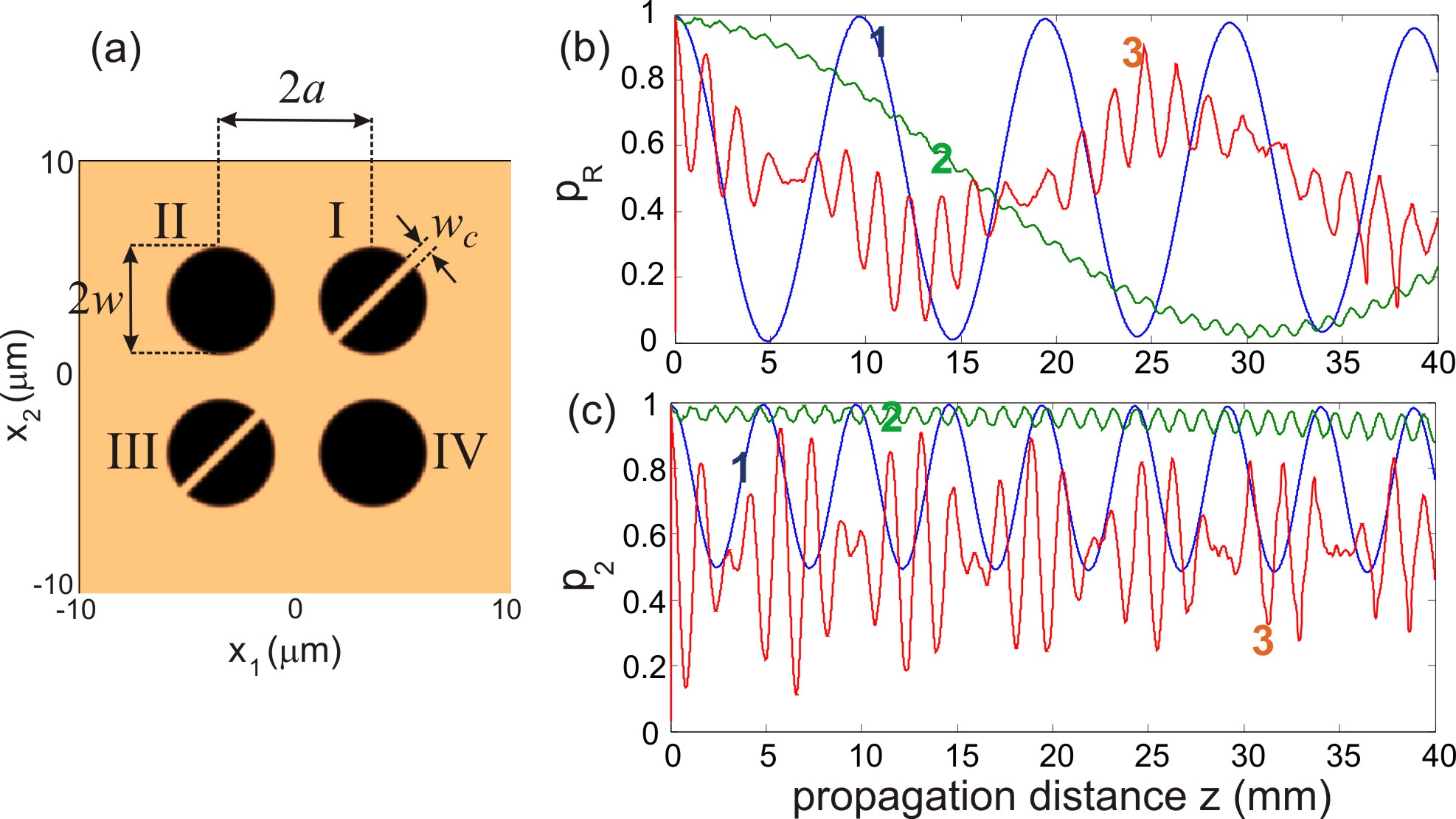}
\caption{ (color online) Light tunneling dynamics in a four-core
fiber. (a) Profile of the optical potential $n_s-n(x_1,x_2)$ (core
diameter $2w= 5 \; \mu$m, core spacing $2a=7 \; \mu$m, index change
$\Delta n=0.005$). (b) and (c) show the evolution of $p_R$ and
$p_2$, respectively, for increasing values of the width $w_c$ of the
cut in guides I and III. Curve 1: $w_c=0$ (non-interacting bosons);
curve 2: $w_c= 0.6 \; \mu$m (correlated pair tunneling); curve 3:
$w_c=  2 \; \mu$m (tunneling of a fragmented pair).}
\end{figure}
Note that the behavior of both the percentage of bosons in the right
well and the same-site boson probability reproduces very well the
different tunneling regimes previously found in Fig.2. \par The good
description of the tunneling dynamics offered by the coupled-mode
equations (5) indicates that the tunneling dynamics of two bosons,
shown in Fig.2, is rather insensitive to the specific shapes of the
guides, and could be thus observed in simpler optical structures.
For example, in Fig.4 it is shown that a similar dynamical behavior
can be realized using a microstructured optical fiber with four
circular cores of radius $w$ and step-index $\Delta n$, in which a
cut with variable width $w_c$ is applied to the cores I and III to
mimic boson repulsion [Fig.4(a)]. As the cut width $w_c$ (i.e. the
interaction strength) is increased, a transition from independent
Rabi oscillations (curve 1) to correlated pair tunneling (curve 2)
and to tunneling of a fragmented pair (curve 3) ic clearly observed.

\section{Conclusions}
In conclusion, an optical realization of the tunneling
dynamics of two interacting bosons in a double-well potential, based
on light transport in a four-core microstructured fiber, has been
proposed. The present results indicate that photonic systems could
provide an experimentally accessible test bench to investigate in a
purely classical setting the dynamical aspects embodied in the
physics of strongly-correlated few-particle quantum systems. As
compared to quantum simulators based on the coherent dynamics of
cold atoms or ions trapped in optical lattices \cite{JPBreview}, the
use of a classical optics simulator enables  a direct access to the
evolution of the multiparticle probability density and could provide
a new route to realize other many-body physical models
\cite{Longhiun,Korsch}. For example, the introduction of gain and
loss regions in the optical structure could offer the possibility to
test in the lab the physics of many-body particles within
non-Hermitian $\mathcal{PT}$-symmetric models \cite{Korsch}.

\acknowledgments

 The author acknowledges financial support by the
Italian MIUR (Grant No. PRIN-2008-YCAAK project "Analogie
ottico-quantistiche in strutture fotoniche a guida d'onda").

\end{document}